# Ralph A Alpher, George Antonovich Gamow, and the Prediction of the Cosmic Microwave Background Radiation

Victor S Alpher
*Austin, Texas, USA*

The first prediction of the existence of "relict radiation" or radiation remaining from the "Big Bang" was made in 1948. This derived from the seminal dissertation work of Ralph A. Alpher. He was a doctoral student of George A. Gamow and developed several critical advances in cosmology in late 1946, 1947, and 1948. Alpher developed the ideas of "hot" big bang cosmology to a high degree of physical precision, and was the first to present the idea that radiation, not matter, predominated the early universal adiabatic expansion first suggested by A. Friedmann in the early 1920s. Alpher & Herman predicted the residual relict black-body temperature in 1948 and 1949 at around 5 K. However, to this day, this prediction, and other seminal ideas in big bang cosmology, have often been attributed erroneously to the better-known George A. Gamow. This article reviews some of the more egregious and even farcical errors in the scholarly literature about Ralph A. Alpher and his place in the history of big bang cosmology. Two such errors are that (a) Alpher was a fictive person; or (b) that like the French mathematician Nicolas Bourbaki, Alpher was a "conglomerate" of theoreticians.©Anita Publications. All rights reserved.

## 1 Could another CMBR history be necessary?

This brief history (with some truly humorous elements) of one phase of early work on the "Big Bang" has been made necessary as a result of flagrant and repeated *errors* in the international scholarly literature concerning the independent contributions Ralph A. Alpher to the development of the big bang cosmology during the $20^{th}$ and $21^{st}$ centuries. George A. Gamow (Г. А. Гамов) was Ralph A. Alpher's thesis advisor during the 1940s at The George Washington University in Washington, D C. Ralph A. Alpher's independent work is often credited to the well-known Russian-Soviet physicist (see Figs 1, 2, & 3). As is demonstrated here, R. Alpher's contribution to cosmology was not only the prediction of the CMBR, but also theoretical groundwork on the "Big Bang" theory, and early nucleosynthesis.

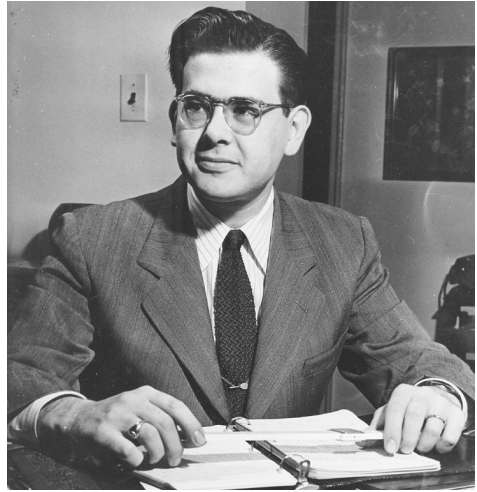

Fig. 1. Photograph of Ralph A Alpher, Ph D, approximately 1950. Copyright, Estate of Ralph A Alpher, Victor S Alpher, Executor (Alpher Papers).

*Corresponding author :*
e-mail: alphervs@gmail.com (Victor S Alpher)

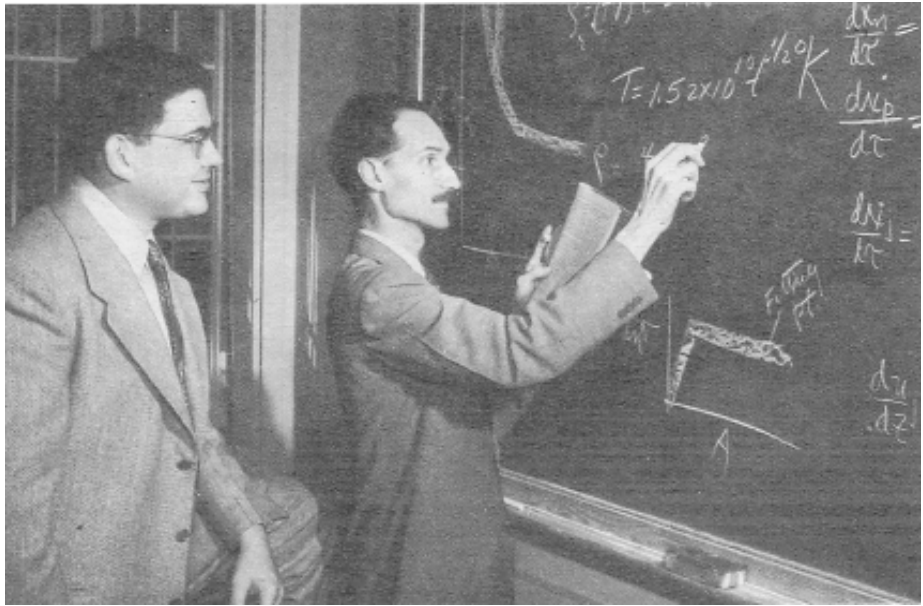

Fig. 2. Ralph A. Alpher and Robert C. Herman working at the blackboard, late 1940s. (Alpher Papers)

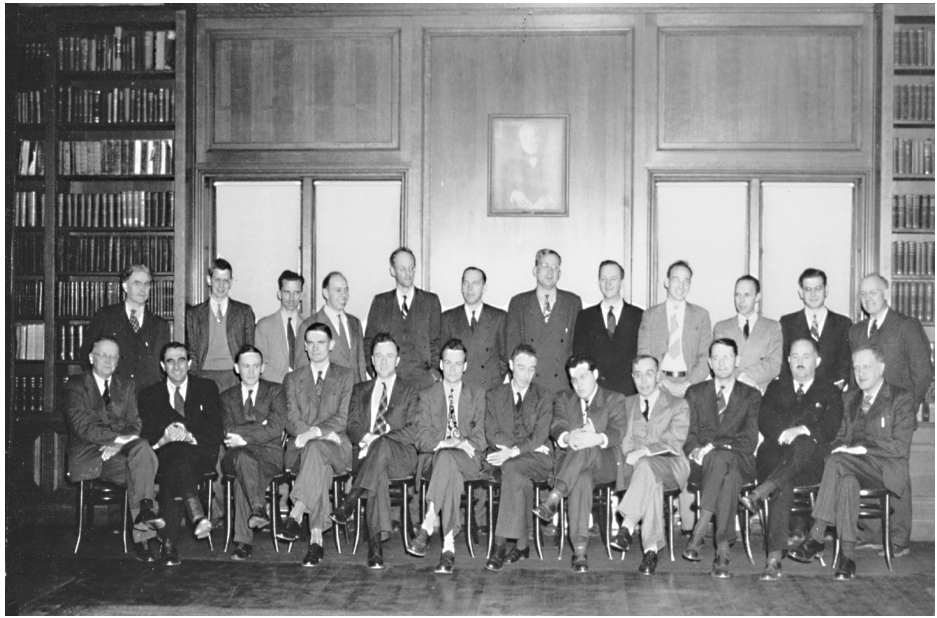

Fig. 3. 1947 Meeting of the Washington Conference on Theoretical Physics, R A Alpher is standing, second from the right in the back row; George A. Gamow is the sixth from the right, standing (Alpher Papers).

Communication about cosmology from the American and Western-European scientific community and the Soviet scientific community has long been strained. It became more so after Gamow left the Soviet Union permanently after a year in France, in 1934 (Gamow, 1970) . He settled in as a Professor of Physics at The George Washington University in Washington, D.C., unable to find a suitable position in Europe.

George Gamow became very well known in the United States (see, e.g., Orndorff, 2013). Through writing popular books of physics and cosmology, he had a large fan base in the "baby boom" generation following World War II.

The country was particularly enamored with physics and physicists following World War II and the detonation of the atomic bombs over the Japanese cities of Hiroshima and Nagasaki. One of Gamow's series of popular books on physics and nuclear science created a quizzical figure known as "Mr. Tompkins.". The series had an important influence on a new generation of fledgling scientists growing up in the 1950s (personal communication, Robert W. Wilson, July 27, 2007). This led to the award of the U N's Kalinga Prize in 1956. This supported a lecture and speaking tour of Asia and the other countries in the southern hemisphere, increasing his popularity.

Gamow wrote of taking a course in Relativity from A. Friedmann (Фридман; his name has been translated with two "n"s; this is a convention in translation from Russian to English). However, because of Friedmann's untimely and early death in 1925, Gamow finished his doctoral-level work with Professor Kruitkov, although never formally obtaining a doctorate. His experience with Kruitkov did not live up to his expectations and dreams of working on relativistic astrophysics with Friedmann (Gamow, 1970, p. 44). Many cosmologists trace the beginning of the idea of an expanding universe to Friedmann (Alpher & Herman, 2001).

**2 Alpher, Gamow, and the CMBR Prediction**

I became intimately familiar with the work of Alpher, Herman, and Gamow from the point at which I was able to read. By the time the Cosmic Microwave Background Radiation (CMBR) was "accidentally" measured by Arno Penzias and Robert Wilson at the Bell Telephone Laboratory (Crawford Hill Radiotelescope, Holmdel, New Jersey, U.S.A.) in 1964, I was already aware of its enormous significance to cosmology. This was heralded in a paper in the *Astrophysical Journal* in 1965 (but not by Penzias and Wilson, who only reported the observation itself at 1080 Mc/s). They did not provide a cosmological interpretation of the measurement as evidence of relict "Big Bang" microwave black-body radiation (CMBR). "Interpretation" of their finding was found in a companion piece by R H Dicke *et al* (1965) that turned out to be wanting in many respects. No reference to the previous work of Alpher and Herman in big bang cosmology in 1948 and 1949 was made. George Gamow did publish an estimate of the CMBR in 1953 in association with his election to the Danish Academy of Sciences (Gamow, 1953)—however, he was never a strong or vocal supporter. During the first crucial years following Alpher and Herman's publications, he opposed the idea on theoretical grounds (Alpher, 2012).

During my own scientific training, I became aware especially of the importance of scholarship in the progress of science. Many would probably argue my position to be politically naïve, particularly considering recent revelations concerning the validity of "man-made" global warming (Biello, 2009). In 2013, after 15 years of no appreciable change in global temperatures, scientists have concluded that "global warming" is "not incontrovertible" and what there is, is not anthropogenic (Rose, 2013).

The reason for this is that his independent scientific work was often been misattributed to other scientists – to Gamow, to Herman, to Hans Bethe. My father became keenly aware that Gamow's reputation could eclipse his own work, that of a mere graduate student (V. S. Alpher, 2009). Ralph Alpher referred to a well-known article in the weekly journal ***Science*** concerning this "Matthew effect" (Merton, 1968; see also Jin, Jones, & Lu, 2013 whose study of a "reverse" Matthew effect seems even to suggest that the most prominent in a group of scientists is afforded the best protection from later criticism) from which he drew his main understanding of neglect of his work. R. Alpher and Bob Herman often sought reasons for the misattribution of their collaborative work to Gamow, a discussion that I personally witnessed several times in the 1980s and 1990s. They had too much respect for Gamow, however, to challenge him directly about his neglect of their work in his citations before his death.

Many errors in the recent Russian literature can be traced to the work of A. D. Chernin (1994a, 1994b, 1995) and I. Tkachev at CERN. From Chernin, we might conclude that Gamow's most significant contribution in 20th century science was his work on the Hydrogen Bomb (Chernin, 1994b) Chernin (Tropp, Frenkel, Chernin & Dron, 1993) even wrote that Robert C. Herman was one of Gamow's *graduate students!* John C. Mather repeated this error in the printed version of his Nobel Prize acceptance speech, but also adheres to the traditional nomenclature in his writings, i.e. CMBR for Cosmic Microwave Background Radiation (Mather, 2007; Mather et al., 2013). In fact, Herman's doctorate was earned with a committee loosely "chaired" by H. P. Robertson at Princeton University in 1940. By the time he began collaborative work with Alpher and Gamow in 1948, he was eight years post-doctoral. However, the misattributions and lack of citation is not limited to the Russian literature (e.g. Kirilova & Chizhov, 1998).

All *three* men knew each other very well during World War II – each worked on Secret military ordnance work for the U.S. Navy (variously at the Naval Research Laboratory, the Naval Ordnance Laboratory, or the Applied Physics Laboratory of Johns Hopkins University at 8621 Georgia Avenue in Silver Spring, Maryland (Alpher, 2008a)). Albert Einstein also worked on contract to the U.S. Navy Bureau of Ordnance and High Explosives as did Gamow. Although Einstein lived in Princeton, New Jersey, he was regularly briefed in person by Gamow who carried materials by train from Washington to Princeton and back. When finally approved at the appropriate security level in 1948, Gamow made many trips to Los Alamos.

A rather more astounding attribution concerning Ralph A. Alpher was repeated by Bamberg (2002). In referring to a well-known paper by Alpher, Bethe, and Gamow (1948) he refers to an article by Halmos (1957) and states that Halmos' asserted that the "fictitious 'first author'" was invented by his *famous co-authors* (Bamberg, 2002, p. 1429; see also Clark, 2005). The implication is presumably that George A. Gamow and Hans Bethe would have invented "Ralph A. Alpher" for some purpose.

I could produce my father's birth certificate (which we know is no guarantee (Corsi, 2011; Vuoto, 2013) – but this is most amusing. He was extensively vetted by the Federal Bureau of Investigation (FBI; internal police investigation force which investigates the backgrounds and associates of persons who will be exposed to information vital to national security) to retain his Q clearance (the highest level of security clearance) in 1952. In fact, for much of his early career, R. Alpher had a higher security clearance than did Gamow—formerly a Red Army Artillery Officer. Gamow did have some prescience to include much of Alpher's work on the "Origin of the Chemical Elements" as an Appendix in the 3$^{rd}$ edition of *Structure of the Atomic Nucleus and Nuclear Transformations* (Gamow and Critchfield, 1949). He was again subjected to an F.B.I. investigation prior to the award of the National Medal of Science.

Professor Tkachev has continued to perpetuate the erroneous attribution of the theoretical prediction of the CMBR at about 5K to *Prof Gamow* (Tkachev, 2004; Tinyakov & Tkachev, 2006). This is incorrect. The prediction was *first* made by Ralph A. Alpher and Robert C. Herman in 1948 (Alpher & Herman, 1948, 1949); the prediction was based upon Ralph A. Alpher's work on nucleosynthesis, conducted while he was one of Gamow's graduate students (Alpher, 2009).

George A. Gamow, who was well known for making practical jokes (**грубая шутка**), included placing winner of the 1937 Nobel Prize Hans Bethe on R. Alpher's dissertation defense committee. Bethe's name was included on the major publication about the origin of the elements and the neutron-capture theory, submitted to *Physical Review Letters* on February 18, 1948—months before the dissertation defense. The ethics of Gamow's actions are certainly questionable and Alpher objected, but Gamow wanted to get the material into the literature as soon as possible--a habit he had already developed.

Consequently, more than 300 persons, including representatives of the national and international press, attended Ralph A. Alpher's public defense of his dissertation on the *formation of the chemical elements* in a hot, dense "Big Bang." None of the press reports indicated Alpher's absence at his own defense. Professor Gamow did help to complicate the history with a subsequent estimation of the CMBR in 1953, although he had not supported the idea for several years. His own mathematical work was rough and to some, amateurish. In some cases, he did not refer to the work of Alpher. Of Alpher and Herman, he tried to make some amends in 1967 with a co-authored paper, but this did not disambiguate the origin of his ideas from those of Ralph Alpher. Herman was ill after having had a heart attack prior to launch of this paper, and did not write a word of it. Following the observation of the CMBR by Arno Penzias and Robert Wilson (1965), the interpretation by Dicke, Peebles, Roll, and Wilkinson (1965) gave no credit to Alpher, Herman, or Gamow's years of work on the problem of the early expanding Universe and the origin of the chemical elements.. They also did not provide references to other observations, many of which were made by Soviet physicists and radioastronomers (see Novikov, 1984; Trimble, 2006).

As mentioned above, Gamow did try to rectify the situation, somewhat. After the "discovery" of the CMBR by Arno Penzias and Robert Wilson (Penzias and Wilson, 1965), which received international recognition, Gamow published a paper shortly before his death in collaboration with Alpher and Herman (Alpher, Gamow, & Herman, 1967). What could this be be but too little, too late? He mixed his interest in *protogalaxies* with Alpher and Herman's work, only further confusing anyone interested in the history of their ideas. Alpher did his best to help Gamow with the mathematical underpinnings of new ideas on galaxy formation (Alpher, under review). Gamow's numerous publications during the 1950s and 1960s are open to a variety of interpretations. Nevertheless, in the final analysis, they caused the umbra of his professional and personal shadow to block out historical memory in the physics community such that someone receiving their doctorate in the 1960s could say they were unaware of Ralph A. Alpher's pioneering work, not have a strong motivation to do any background library research (Alpher, 2012). Others would later make light of it, although the members of the National Academy of Sciences (Neil deGrasse Tyson, personal communication 26 July 2007) made certain his work was not forgotten (see the citation in the Fig. 4 at the end of this paper).

Another oddity of this history is that Alpher, in his dissertation, first proposed that *radiation dominated the early universe, rather than matter* (Alpher, 1948):

"If the cosmological model leading to Eq. (37) is believed, then the starting time of the process should have been several orders of magnitude earlier than a $10^7$ seconds. In addition, we have seen that the temperature obtaining during the process must have been of the order of $10^5$ eV $\cong 10^9$ K. But the density of black body radiation at this temperature would have been

$P_{\text{radiation}} = 0.848 \times 10^{-35} \, T^4 \cong 10 \text{ gm/cm}^3$ [handwritten in India ink]

This is many orders of magnitude greater than the density of matter given by the particular cosmological model used. It would appear, then, *that radiation was dominant* [italics added VSA] in determining the behavior of the universe in the early stages of its expansion, and the cosmological model which we have used is not correct. An interpretation of the starting time and initial density for the neutron-capture process will therefore require the development of a new cosmological model." (RAA's Dissertation, p. 66). R. Alpher found this important enough to repeat in the Appendix to his dissertation:

On p. 16 of the Appendix to the dissertation RAA writes:

*"As has been pointed out, the temperature during the capture process must have been of the order of $10^5$ eV $\cong 10^9$ K. The density of the radiation at such a temperature is of the order of 1 gm/cm$^3$, many orders of magnitude higher than the initial density given by Eq. 23). It is suggested, therefore, that radiation controlled the initial stages of the universal expansion, and that the cosmological model leading to Eq. (8) can not be correct. The initial density and starting time for the neutron-capture process must therefore be reinterpreted in terms of a different cosmological model…..We may conclude that wheras temperature was the controlling factor for the start of the process, temperature decrease in the expansion did not terminate the process."*

On p. 18 of the Appendix to his dissertation RAA concludes:

> *"Because of the adiabatic expansion, temperatures were still quite high after the neutron-capture process ceased, high enough in fact to support thermonuclear reactions of these elements with protons at a very accelerated rate. It is suggested that the universe went through a state of element formation to a state when only thermonuclear reactions were possible, and, with further expansion, to a state when no nuclear reactions at all went on. The growth of stellar configurations somehow began after this stage.*
>
> *"The theory presented for the formation and relative abundance of the elements is obviously in preliminary form. To effect significant improvement in the theory requires that the β-decay of neutrons and a correct cosmological model be introduced into the mathematical formulation of the process. In addition, a more detailed knowledge is requires of the capture cross sections of the elements, particularly the light elements."*

These turned out to be ideas of inestimable importance to the growth of cosmology. Yet, the ideas are continually attributed to Gamow (e.g. Kragh, 2005, p. 180). However, Gamow published no such formulations prior to 1948 (e.g. Gamow, 1946a, 1946b). From 1948 onward, Gamow was dependent on his former student Alpher to provide him lecture materials and slides so that he could give talks about the "Origin of the Elements" He did publish a short independent paper using some of these ideas—when Alpher and Herman saw a second paper, they sent him mathematical corrections by Telegram! (Gamow, 1948a, 1948b)

Gamow wrote Alpher often asking for new drafts of papers on the subject, and slides to use in presentations at locations such as Los Alamos National Laboratory (Alpher, under review). Gamow was a dedicated traveler with substantial funding; Alpher was not able to do so, just having obtained his doctorate in 1948. and having two children, in 1950 and 1954. These talks also gave the impression that this large body of work was Gamow's, whereas Alpher had done the pioneering and careful mathematical work to develop big bang cosmology in 1946-1949. These are the years in which R Alpher developed his dissertation on the origin of the chemical elements in the big bang, and the idea and first temperature calculations of the relict Cosmic Microwave Background Radiation (see also Harper, 2001 on Gamow's research style).

**3 Three Pillars of Big Bang Cosmology: What Next?**

The pre-publication of some of Alpher's dissertation promoted by Gamow (Alpher, Bethe, and Gamow, 1948) simply further clouded the history of ideas of big bang cosmology as it emerged as a true branch of physics. The paper became known as the "α, β, γ" which in one way made it easier to recall, but futher associated the ideas with Gamow, not their actual origins (Alpher, 2012). However, had it become known as "Alpher *et al*, 1948" we might never have heard of it again…..it is difficult to reconcile. It is NOT simply an issue of order of authorship.

We know that some would argue that these details of the historical record of scholarship in science are of interest – but only to historians of science. I would disagree, because they are critical to the trust the public places in the integrity of science, its methods, and its hypotheses. Today, such debate is closely tied to our understanding of climatology; the methods of research on climate going back tens of thousands of years are coming under close scrutiny (Monckton, 2008). They should be. "Consensus" is difficult to achieve, even in data-based climatology; one Nobel Prize recipient (1973; Ivar Giaever) *resigned* from the American Physical Society (APS) over the APS "conclusive" position statement on "anthropogenic global warming." Giaever is not the only individual to have difficulty with grand sweeping "conclusive science" (see Strassel, 2009). Fortunately, at least until the Cosmic Microwave Background Explorer (COBE) satellite was conceived, cosmological research was isolated from all-important government funding. R. Alpher proposed some of the types of experiments possible from a space platform such as COBE provided as early as 1962 when the General Electric Company was making such proposals for government-funded research (Alpher, 2012). Such funding often distorts a research agenda—in the case of COBE, all the way to two more Nobel Prizes in Physics, 2006. The methods of research on cosmology and particle astrophysics are no doubt of equal importance, as is the history of ideas in each field. Without them, such projects would fail on the platform as surely as the early A-2 (V-2, Germany's Vengeance Weapon) on the platform at Pennemünde.

**4 Do scholarship and history matter?**

"Forgetting" earlier work on the CMBR is not a problem unique to Russian or Soviet cosmology and astrophysics. B Parker noted this in his 1993 work: "Dicke admitted later that it was an oversight. He had completely forgotten about Gamow and his students' work. Furthermore, Peebles, in making his calculations, had not checked the literature. *He just assumed that no one had worked on the problem*" (Parker, 1993, pp. 120-121, italics added). Actually, there is a bit more to Peebles' story, considering his approach to researching a new problems was to start on his own, and not to check the history or scholarly literature (Alpher, 2012). Russian scientists will probably find this degree of lassitude even among American scientists somewhat disheartening, as learning English and gaining access to English-language astrophysics and physics journals is essential to their work. However, Parker's findings are consistent with the oral histories available from the Neils Bohr Library of the American Institute of Physics (see Alpher, 2012).[1*]

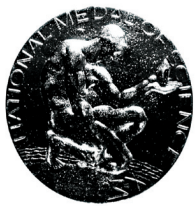

Fig. 4. Certificate from the National Medal of Science, awarded to Ralph A. Alpher, 2005. Citation: The National Medal of Science is awarded by The President of the United States of America to Ralph A. Alpher "For his unprecedented work in the areas of nucleosynthesis, for the prediction that the universe expansion leaves behind background radiation, and for providing the model for the Big Bang theory."  The recognition came very late in his lifetime, but it stands as a beginning.

---

[1*] Martin O Harwit has a unique perspective, as he conducted the interviews of Alpher, Herman, and Peebles for the Oral History Project of the Niels Bohr Library at the American Institute of Physics. In a letter to V S. Alpher dated December 22, 2012, Harwit wrote, commenting on Alpher (2012): "It may be fun to work something out for yourself (referring to P.J.E. Peebles), but that does not justify denying priority to those who did it first." Peebles insists (personal communication by e-mail, dated September 29, 2013) that any historical analysis is subject to the "hazards of recollections" which presumably make any analyses that take into account any later recollections suspect, and methodologically unsound. It may be convenient to ignore history, but it is also tendentious and misleading, in this author's experience. Ignoring history may appear like a clean approach, but then why do we study the history of ideas at all?

(Alpher Papers)

Nevertheless, the National Science Foundation committee tasked with making recommendations put forward the award of the U S National Medal of Science to Ralph A Alpher, 2005 (award was made on 27 July 2007). The certificate is reproduced in Fig. 4.

This award was made only weeks before his passing on 12 April 2007—but soon enough to give Ralph A Alpher some vindication for decades of patience and dedication to cosmology and theoretical physics. The answer to the question posed by the heading of this section—"Do Scholarship and History Matter?" ultimately can be answered positively, especially to those who take them seriously. From Hillel we read in Ethics of the Fathers (1:14), "If I am not for myself, who will be for me? But if I am only for myself, who am I? If not now, when?" In science, self-promotion is not ultimately the way to go. The history of ideas and concepts in science shows that science does not proceed in the orderly manner as usually presented to undergraduates.

I did not hear Ralph A Alpher utter this, but he surely learned the principle at a young age—it is an effective ethical guideline.

**Acknowledgments**

I extend special appreciation to Virginia Trimble, Dwight Neuenschwander, Roger Stuewer, Stephen G. Brush, John C. Mather, Neil deGrasse Tyson, Stephen Bennett, Miguel Saldaña, Phil Kosky, Mark Turner, Robert "Bev" Orndorff, Howard and Sandra Raben, P. James E. Peebles, Robert Wilson, Ron Humphreys, Bob Bodin, Tony Johnson, Kevin Miller, Neil Blumhofe, Joy and David Wharton, Kristina Pavlova, Julie Marie Fontaine, Arnt "Bill" Erickson. Ken Gutshall, Mary Jane Hammerle, and Eamon Harper for their continued interest in and support of my scholarship. Somewhere there are folks who really know what motivates a scholar who works on a lean schedule of variable reinforcement.

The Librarians and Staff of the Kuehne Mathematics-Physics-Astronomy Library of the University of Texas Libraries have been always helpful: Molly White and David Gilson.

## Appendices

**Appendix 1.** Космологи Часто Ошибаются Но Никогда Не Сомневаются, "Cosmologists are Often Wrong but Never in Doubt." Lev Landau (a facsimile that Ralph A Alpher kept tacked to a bulletin board in his office).

**Appendix 2**. "Alpher Papers" refer to documents and photographs in the Personal Papers of Ralph A. Alpher, Victor S. Alpher, Archivist. All material copyright © 2013.

**Appendix 3**. An early version of this paper was deposited in unpublished form at <physics-online.ru>.